\documentstyle[epsf]{article}

\newcommand{\be}{\begin{equation}}
\newcommand{\ee}{\end{equation}}

\def\Journal#1#2#3#4{{\em #1} {\bf #2}, #4 (#3)}

\def\PRD{{\em Phys. Rev.} D}

\begin{document}

\begin{flushright}
Liverpool Preprint: LTH 550\\
 \end{flushright}
  
\vspace{5mm} \begin{center} {\LARGE \bf Exotics
 \footnote{Prepared for {\em World Scientific Publishing Company}
(International Review of Nuclear Physics, Vol. 9, Hadronic Physics from
Lattice QCD, edited by A. M. Green)}}\\[10mm]

 {\bf  C.~Michael \\
Theoretical Physics Division, Dept. of Mathematical Sciences, 
          University of Liverpool, Liverpool L69 3BX, UK 
 }\\[2mm]

\end{center}

\begin{abstract}

We review lattice QCD results for glueballs
(including  a discussion of mixing with scalar mesons), hybrid mesons
and other exotic states (such  as $B_s B_s$ molecules).
\end{abstract}

\section{Introduction}     
 \label{ex.sect1}

 Quantum Chromodynamics has emerged as the unique theory to 
describe hadronic physics. It is formulated in terms of gluonic and quark
fields. The only free parameters are the scale of the coupling (usually
called $\Lambda_{QCD}$)  and the quark masses defined at some
conventional energy scale.

 Where large momentum transfers occur,  the  effective coupling becomes
weak and a perturbative treatment is valid: in  this domain the theory
has been tested directly by experiment. However, because the effective
coupling is weak for these processes that can be  described by
perturbation theory, they are necessarily not the dominant hadronic 
processes. A typical hadronic process will involve small momentum 
transfers and so has to be treated non-perturbatively.

 In this non-perturbative r\'egime, the description of hadrons is quite
far removed  from the description of the gluonic and quark fields in the
QCD Lagrangian. Because only colour-singlet states survive, the hadrons
are  all composites of quarks and gluons. One example emphasises this:
the  nucleon has a mass which is very much greater than the sum of the 
quark masses of the three valence quarks comprising it. This extra mass 
comes from the gluonic interactions of QCD. Another way to view this is
that  the na\"{\i}ve quark model is a useful phenomenological tool but
has constituent quarks with masses much greater than the QCD masses (ie
masses  as defined in the Lagrangian). It is important to understand why
this  is approximately what QCD requires and to find where QCD departs
from  the na\"{\i}ve quark model.

 One way to characterise the manner in which QCD goes beyond the na\"{\i}ve
 quark model is through the concept of exotic states. Here exotic is
taken to mean  \lq not included in the na\"{\i}ve quark model\rq. 
 In order to discuss exotic states, we need to summarise what the
na\"{\i}ve quark  model contains. Basically the degrees of freedom are the
valence quarks  (ie quark-antiquark for a meson and 3 quarks for a
baryon) with masses and interactions  given by some effective
interaction. The consequences of this are that only certain $J^{PC}$
values  will exist and that the number of states with different quark
flavours is  specified. So, concentrating on mesons made of the three
flavours of light quarks ($u,\ d,\ s$),  one expects a nonet of mesons
with the flavours ($\bar{u} d, \  \bar{d} u,\  \bar{u} u \pm \bar{d} d,\
\bar{s} s,\ \bar{u} s,\ \bar{d} s,\   \bar{s} u,\ \bar{s} d$).  This is
indeed  what is found for vector mesons ($\rho,\ \omega,\ \phi,\ K^{*}$). 
It is also possible within the quark model for the flavour-singlet states 
($\bar{u} u + \bar{d} d,\  \bar{s} s$) to mix, as found for the pseudoscalar 
mesons. What would be exotic is for a tenth state to exist. 
  For mesons with orbital angular momentum $L$  between the quark and
antiquark the  allowed $J^{PC}$ values are shown below.  Thus spin-parity 
combinations such as $0^{--}, 0^{+-}, 1^{-+}, 2^{+-}$ are termed
spin-exotic  since they cannot be made from a quark plus antiquark
alone. 

\begin{table}
\begin{center}
\begin{tabular}{|c|c|ccc|} \hline
  $L$ &   $J^{PC}$ &$J^{PC}$ &$J^{PC}$ &$J^{PC}$ \\ \hline
0 & $0^{-+}$ &          & $1^{--}$ &           \\
1 & $1^{+-}$ & $0^{++}$ & $1^{++}$ & $ 2^{++}$ \\ 
2 & $2^{-+}$ & $1^{--}$ & $2^{--}$ & $ 3^{--}$ \\ 
\hline
\end{tabular}
\end{center}
\end{table}

It has been a considerable challenge to  build a machinery that allows
non-perturbative calculations in QCD with all systematic  errors determined.
 The most controlled  approach to non-perturbative QCD is via
lattice techniques in which space-time is discretized and time is taken
as  Euclidean. The functional integral is then evaluated numerically
using  Monte Carlo techniques. 

  Lattice QCD needs as input the quark masses and an overall scale
(conventionally  given by $\Lambda_{QCD}$). Then any Green function can
be evaluated by taking an average of suitable combinations of the
lattice fields in the vacuum samples. This allows masses to be studied 
easily and matrix elements (particularly those of weak or
electromagnetic currents)  can be extracted straightforwardly.
  Unlike experiment, lattice QCD can vary the quark masses and can also 
explore different boundary conditions and sources. This allows a wide
range of  studies which can be used to diagnose the health of
phenomenological models as well as casting light on experimental data.

One limitation of the  lattice approach  to QCD is  in exploring
hadronic decays because the  lattice, using Euclidean
time, has no concept of asymptotic  states. One feasible strategy is to
evaluate the mixing between states of the same  energy - so giving some
information on on-shell hadronic decay amplitudes.

 There is an interesting theoretical world in which the quark degrees 
of freedom are removed from QCD, leaving pure gluo-dynamics. This is
also  known as pure Yang-Mills theory.  It is a self-consistent  theory
which has the full non-perturbative gluonic interaction. It turns out
that this gluonic interaction does produce the salient features of QCD:
asymptotic freedom, confinement,  etc.  It is of interest to explore
the spectrum  in this case:  the states are called glueballs. 

 It is also of interest to consider the propagation of quarks  in this
gluonic theory. The quarks are treated   as in the Dirac 
equation and they propagate through the  gluonic ground state.  This
approach is known  as the quenched approximation. Again this turns out
to be  a very useful approximation: chiral symmetry breaking occurs for
example. This quenched approximation is in contrast to the  full quantum
field theory of QCD where there would be quark loop effects in the
ground state also. Thus in the quenched approximation there will be
inconsistencies:  the theory is not unitary. However, for heavy quarks
it will be a good approximation since  heavy quark loops are suppressed 
and it may be adequate to describe some features of lighter quarks.  
Moreover, many phenomenological models are appropriate to the quenched case 
and so can be compared with quenched QCD.

\section{Glueballs and scalar mesons}
 \label{ex.sect2}

\subsection{Glueballs in quenched QCD}
 \label{ex.sect2.1}

Glueballs are defined to be hadronic states made primarily from gluons.
The full non-perturbative gluonic interaction is included in quenched
QCD.  A study of the glueball spectrum in quenched QCD  is thus of great
value. This will allow experimental searches to be  guided as well as
providing calibration for models of glueballs. A non-zero glueball mass
in quenched QCD is the  ``mass-gap'' of QCD. To prove this rigourously
is one of the major challenges  of our times. Here we will explore the
situation using computational techniques.

In lattice studies, dimensionless ratios of  quantities are obtained. To
explore the glueball masses $m$, it is appropriate to combine  them with
another very accurately measured quantity to have a dimensionless 
observable. Since the potential between static quarks is very accurately
measured from the lattice, it is now conventional~\cite{sommer} to use
$r_0$ for this comparison.  Here $r_0$ is implicitly defined by $r^2
dV(r)/dr = 1.65$ at $r=r_0$ where $V(r)$ is  the potential energy
between static quarks which is easy to determine accurately  on the
lattice.  Conventionally  $r_0 \approx 0.5$ fm.

\begin{figure}[bt] 
\vspace{-2.0cm}  
\epsfxsize=10cm\epsfbox{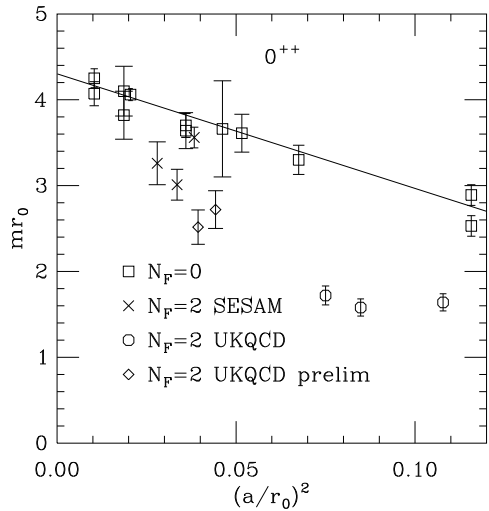}

 \caption{ The value of mass of the  $J^{PC}=0^{++}$  glueball state
from quenched data ($N_F=0$){\protect\cite{DForc,MTgl,ukqcd,gf11}}
in units of $r_0$ where $r_0 \approx 0.5$ fm. The straight line  shows a
 fit describing the  approach to the continuum limit as $a \to 0$.
  Results~{\protect\cite{sesam,cmcm176,cmcm202}}
for the lightest scalar meson  with $N_F=2$ flavours of sea quarks
are also shown.
   }
 \label{ex.fig1}
\end{figure}

 Theoretical analysis  indicates that for  Wilson's discretisation of
the gauge fields in the quenched approximation,  the dimensionless ratio
$mr_0$ will differ from the continuum  limit value by corrections of
order $a^2$.  Thus in fig.~\ref{ex.fig1} the mass of the
$J^{PC}$=$0^{++}$  glueball is plotted versus the lattice spacing $a^2$.
The straight line then shows the continuum limit obtained  by
extrapolating to $a=0$. As can be seen, there is essentially no need for
data  at even smaller $a$-values to further fix the continuum value. The
value shown  corresponds to $m(0^{++})r_0=4.33(5)$.  Since several
lattice groups~\cite{DForc,MTgl,ukqcd,gf11} have measured these 
quantities, it is reassuring to see that the purely lattice observables
are in  excellent agreement. The publicised difference of quoted
$m(0^{++})$ from  UKQCD~\cite{ukqcd} and GF11~\cite{gf11} comes entirely
from relating quenched lattice  measurements to values in GeV.

In the quenched approximation, different hadronic observables differ
from experiment  by factors of up to 10\%. Thus using one quantity or
another to set the scale, gives an overall systematic error.  Here  the
scale is set by taking the conventional value of the string tension
(determined  from potential models and from hadronic lattice studies),
$\sqrt{\sigma}=0.44$ GeV, which then corresponds to $r_0^{-1}=373$ MeV
(or $r_0 = 0.53$ fm). An overall systematic error of 10\% is then to be
included to any  extracted mass. This yields $m(0^{++})=1611(30)(160)$
MeV where the second error is the systematic  scale error. Note that
this is the  glueball mass in the quenched approximation -  in the real
world significant mixing with $q \bar{q}$ states etc may modify this
value substantially, as we discuss below.

\begin{figure}[t]
\epsfxsize=10cm\epsfbox{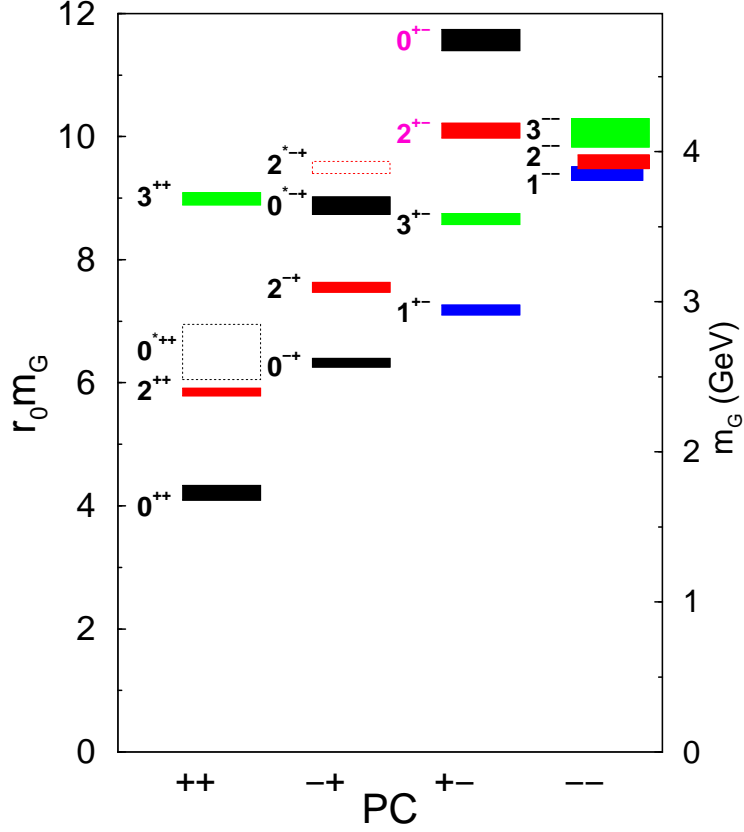}
 \caption{ The continuum glueball spectrum{\protect\cite{mpglue}}. 
 }
 \label{ex.fig2}
\end{figure}

In the Wilson approach, the next lightest glueballs
are~\cite{MTgl,ukqcd} the tensor $m(2^{++})r_0=6.0(6)$  (resulting in  
$m(2^{++})=2232(220)(220)$ MeV) and the pseudoscalar $m(0^{-+})r_0=
6.0(1.0)$. Although the Wilson discretisation provides a definitive
study of the lightest ($0^{++}$)  glueball in the continuum limit, other
methods are competitive for the determination of the mass  of heavier
glueballs.  Namely, using an improved gauge discretisation which has 
even smaller discretisation errors than the $a^2$ dependence of the
Wilson discretisation,  so allowing a relatively coarse lattice spacing
$a$ to be used. To extract mass values, one has to explore the time
dependence of correlators and for this reason,  it is optimum to use a
relatively small time lattice spacing. Thus an asymmetric  lattice
spacing is most appropriate.  The  results~\cite{mpglue}  are shown in
fig.~\ref{ex.fig2} and for low lying states are that
$m(0^{++})r_0=4.21(11)(4)$,  $m(2^{++})r_0=5.85(2)(6)$,
$m(0^{-+})r_0=6.33(7)(6)$ and $m(1^{+-})r_0=7.18(4)(7)$. It will be very
difficult to identify experimentally states corresponding  to these
heavier glueballs since the spectrum is rich in $q \bar{q}$ states of
those  quantum numbers at those  mass values and there will thus be
considerable mixing.

One signal of great interest would be  a glueball with $J^{PC}$ not
allowed for $q \bar{q}$ - a spin-exotic glueball or {\em oddball} -
since it would  not mix with $q \bar{q}$ states. These states are
found~\cite{MTgl,ukqcd,mpglue} to be  high lying: considerably above
$2m(0^{++})$. Thus they are  likely to be in a region around 4 GeV where
it is very difficult to separate states unambiguously by experiment. 

As well as the mass of a glueball, it is possible to study their
physical size. In principle this is determined by measuring the matrix
element  $ \langle G | J | G  \rangle $ where $J$ is some local current
which couples to the glueball.  Since glueballs have no flavour, the
energy-momentum tensor is the most  appropriate choice. A preliminary
study has been made, albeit with large  systematic errors~\cite{tickle}
and finds a radius $0.9\pm 0.3$ fm. This approach should be contrasted
with what has come to be called the Bethe-Saltpeter wavefunction  which
is obtained from $ \langle G|L|0 \rangle $ where $L$ is the lattice
operator used to create a glueball state from  the vacuum. Within a
lattice calculation, it is easy to measure the dependence of this 
overlap on the spatial extent of $L$~\cite{glue:bs}, but difficult to
interpret the result.  

 Related information can be obtained  from a study of the gluelump: the
state with one static colour source in the octet representation  with a
gluonic field making it a colour singlet. This would be of physical 
relevance should a massive gluino exist: it would be the glueballino, a 
gluino-gluon bound state. The spectrum~\cite{fm} and spatial
distribution~\cite{jorysz}  have been studied. 

 Another topic which is of mainly theoretical interest is the glueball 
mass (as a dimensionless ratio to the string tension) as the number of
colours $N_c$ is varied from 3. The SU(2) Yang-Mills  theory has often
been studied, especially as  it is computationally simpler. Recently a
study of SU(N) for N=4 and 5 has been made.  The summary~\cite{ncgb} is
that N=$\infty$ is relatively close to N=3. This  has theoretical
implications since the N=$\infty$ theory is formally simpler. 
 For a comparison of these lattice results with ADS supergravity 
see ref.~\cite{brower}.

\subsection{Scalar mesons in quenched QCD}
 \label{ex.sect2.2}

 In quenched QCD the flavour singlet ($f_0$) and non-singlet ($a_0$) 
scalar mesons are degenerate. In full QCD this degeneracy is split  by
disconnected quark diagrams but these are omitted from the quenched
approximation. This same feature of the quenched approximation implies
that the  $\eta$ meson is wrongly treated - it will be degenerate with
the $\pi$. This implies that the scalar meson propagation can have the
wrong sign~\cite{fnal} because the $\eta \pi$  intermediate state is
mistreated (once quark loops are allowed in the vacuum then this 
anomaly is  removed). For light quarks of mass corresponding to the
strange quark or heavier, it is expected that this  anomaly is
relatively unimportant.  Thus the  measurement of the mass of the  $q
\bar{q}$ scalar meson can be particularly  unreliable in the quenched
approximation.

Even though the mixing of the glueball and  $q \bar{q}$ states is not
implemented  in the quenched approximation, one can determine the mixing
matrix element. This can then be used to estimate the result of the
mixing by  hand (by using a mass matrix for example). On a rather coarse
lattice ($a^{-1}  \approx 1.2$ GeV), two groups  have attempted to
measure  this mixing~\cite{weinssg,cmcm176}. Their results expressed as
the mixing for two degenerate quarks of mass around the strange quark
mass  are similar, namely $E \approx 0.3$ GeV~\cite{weinssg} and 0.5
GeV~\cite{cmcm176}.  This is a relatively large mixing (if the glueball
and scalar meson states  were degenerate they would be split by $\pm E$).

 An exploratory attempt to extrapolate this mixing to  the
continuum~\cite{weinssg} gave a very small mixing of 61(45) MeV, while
the  other determination~\cite{cmcm176} uses clover improvement so order
$a$ effects in the extrapolation to the continuum are suppressed and one
would not expect a significant decrease in going to  the continuum
limit. 
 What this discussion shows is that precision studies of the mixing on 
a quenched lattice have not yet been achieved. Furthermore the problems
with the  scalar meson propagation in the quenched approximation
discussed above also limit progress.

As well as this mixing of the glueball with $q \bar{q}$ states, there
will be  mixing  with $q \bar{q} q \bar{q}$ states which will be
responsible for the  hadronic decays. A first attempt to study
this~\cite{gdecay}  at a coarse lattice spacing yields an estimated
width for decay to two pseudoscalar mesons from the scalar glueball of
order 100 MeV.  A more realistic study  would involve taking account of
mixing with the $n \bar{n}$ and $s \bar{s}$ scalar mesons as  well.

\subsection{Scalar mesons in full QCD}
 \label{ex.sect2.3}

 It is now feasible to explore the flavour-singlet scalar meson spectrum
including the quark loops in the vacuum, ie in full QCD.  From dynamical
fermion studies with $N_f=2$, one can determine the  flavour singlet and
non-singlet mass spectrum.  What is found~\cite{cmcm176,cmcm202} is that
the lightest flavour-singlet scalar  meson ($f_0$) is lighter than the
lightest flavour non-singlet ($a_0$).

The interpretation of this study is hampered by the same issue that
hampers the interpretation of experimental data, namely, the  mass
eigenstates are not distinguished as \lq glueball\rq\ or as \lq
quark-antiquark\rq. What one can do is explore the  output spectrum and
deduce what mixing might have occurred. To give an example, where we
restrict here to $N_f=2$  flavours of degenerate quarks, the $f_0$
masses will be $m_0$ and ${m_0}'$ where the  latter is the first excited
state and the flavour non-singlet $a_0$ mass will be  $m_1$. Results for
$m_0$ are given in fig.~\ref{ex.fig1}.  Then one would expect in a
simple $2 \times $2 mixing scenario (ie glueball and $q \bar{q}$ meson)
a mass matrix 

\begin{center}
\begin{math} 
 \left( \begin{array}[h]{cc}
    m_G  &    E \\
    E  &   m_1 \\
\end{array} \right)
\end{math}
 \end{center}

\noindent where $m_G$ is the glueball mass and $E$ the mixing matrix
element. This will have  two mass eigenstates which can be identified
with $m_0$ and ${m_0}'$ so determining the  two free parameters in the
matrix. This approach  explains what is going on but obtains  two
numbers with two parameters, so there is no cross-check.

 One can directly address the issue of the mass of the lightest scalar
singlet meson  from the lattice with $N_f=2$. It is advantageous to use 
as full a basis of lattice operators as possible, including Wilson loops
and  quark-antiquark loops. Including the latter can in practice lead to
a lower value  the ground state  scalar meson mass - see
refs.\cite{ht,cmcm202}.
 Most studies have shown no significant change of the scalar glueball
mass as dynamical quarks are included~\cite{sesam}.  However the larger
lattice spacing result~\cite{cmcm176} shows a significant  reduction in
the lightest scalar mass, as shown in fig.~\ref{ex.fig1}.
 Before concluding that this implies a lower scalar mass in  the
continuum limit, one needs to check whether  an enhanced order $a^2$ 
correction might be present. The origin of the large coefficient of
$a^2$  in quenched glueball studies is usually ascribed to the presence
of  a critical point in the fundamental-adjoint coupling plane  which is
close by in the  usual Wilson approach with zero adjoint coupling. The
extent to which this will be enhanced/reduced when dynamical quarks are
introduced is not clear.  Studies using the same approach  at a finer
lattice spacing~\cite{ht,cmcm202} do suggest that this  large order
$a^2$ effect is significant for dynamical quarks, but studies even
nearer to the continuum  or with improved actions are needed to resolve
this fully.

 A further complication is that as the quark mass is reduced towards the
physical light quark mass,  the decay to $\pi \pi$ becomes energetically
allowed. The study of unstable  particles is a difficult problem in a
Euclidean time formalism~\cite{cmdecay}.  We return to this topic later.

\subsection{Experimental evidence for scalar mesons}
 \label{ex.sect2.4}
 
 In full QCD,  for the favour-singlet states of any given $J^{PC}$, 
there will be mixing between the  $s \bar{s}$ state, the  $u \bar{u}+d
\bar{d}$ state  and the glueball as well as with multi-meson channels.
  It may indeed turn out that no scalar meson in the physical spectrum
is primarily a glueball - all states are  mixtures of glue,  $q
\bar{q}$, $q \bar{q} q \bar{q}$, etc.

 To help with understanding the experimental situation~\cite{pdg}, we
first discuss the  flavour non-singlet states, the $a_0$ with isospin
1. The observed states are  at 980 and 1450 MeV. The lighter state has
dynamics which appears to be closely associated with the  $K \bar{K}$
threshold. The heavier state is not yet very well established but seems
to be a candidate  for a state mostly comprised of $q \bar{q}$, while
the lighter state would  be $q \bar{q} q \bar{q}$. 

 The flavour singlet states ($f_0$) are more numerous. There is a very
broad enhancement  in the $\pi \pi$ S-wave phase shift around 700 MeV
(sometimes called the  $\sigma$), there is a state near the $K \bar{K}$
threshold at 980 MeV  and there are more states at 1370, 1500 and 1710
MeV. Again assuming that the state at 980 MeV is predominantly  $q
\bar{q} q \bar{q}$, this suggests that the three states in the 1300-1750
MeV  energy range are admixtures of the glueball,   $u \bar{u}+d
\bar{d}$ and  $s \bar{s}$. The fact that there are indeed three states
in this energy  region close to the quenched glueball mass of 1600 MeV
is the  strongest evidence for the presence of a glueball.   This has
led to several phenomenological attempts~\cite{weinssg,mixing}  to
describe these three observed states in terms of the lattice input.

 As we emphasised above, in full QCD on a lattice one just obtains 
values for the $a_0$ and $f_0$ masses. In the simplified case of $N_f=2$
flavours of degenerate quark, one does indeed  find~\cite{cmcm202} two
$f_0$ states, and they can be interpreted as mixtures of the  $q
\bar{q}$ and glueball states with the $q \bar{q}$ state having the
properties  found for the $a_0$. 

One useful lattice input would be a determination  of the $a_0$ mass as
the quark mass is varied in full QCD, especially because of the problems
of determining  the $a_0$ mass in the quenched approximation.  At
present, the  full QCD studies~\cite{cmcm176,cmcm202}  are limited to
relatively coarse lattice spacing, so the continuum  limit is not close.
Furthermore, as the quark mass is reduced the  $a_0$ can decay (to $\pi
\eta$) and this will influence the lattice analysis~\cite{milc_a0}.

 \section{Hybrid Mesons}
 \label{ex.sect3}

 A hybrid meson is a meson in which the gluonic degrees of freedom are 
excited non-trivially. The most direct sign of this would be  a
spin-exotic meson, since that could not be created from  a $q \bar{q}$
state with unexcited glue. A spin-exotic meson  could, however,  be  a
$q \bar{q}q \bar{q}$ or meson-meson state and that possibility will be
discussed.
 We first discuss hybrid mesons with static heavy quarks where the
description  can be thought of as an excited colour string.  The
situation  concerning light quark hybrid mesons is then summarised 

\subsection{Heavy quark hybrid mesons}
 \label{ex.sect3.1}

\begin{figure}[bt] 
\epsfxsize=8cm\epsfbox{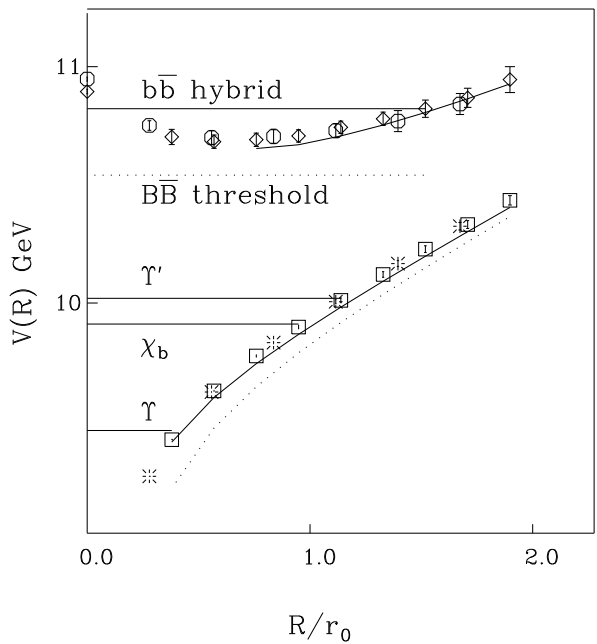}

 \caption{ The potential energy between static quarks at separation $R$
(in units of  $r_0 \approx 0.5$ fm)~{\protect\cite{pm}}. The symmetric
gluonic field  configuration is shown by the lower points while the
$\Pi_u$ excited gluonic configuration is shown above. The energy levels
in these potentials for  $b$ quarks are shown using the adiabatic
approximation. 
   }
\end{figure}

Consider $Q \bar{Q}$ states with static quarks  in which the gluonic
contribution may be excited. We  classify the gluonic fields according
to the symmetries of the system.  This discussion is very similar to the
description of electron wave functions in  diatomic molecules. The
symmetries are  (i) rotation around the separation axis $z$ with
representations labelled by $J_z$ (ii) CP with representations labelled
by $g(+)$ and $u(-)$ and (iii) C$\cal{R}$. Here  C interchanges $Q$ and
$\bar{Q}$, P is parity and $\cal{R}$ is a rotation  of $180^0$ about the
mid-point around the $y$ axis. The C$\cal{R}$ operation is only relevant
 to classify states with $J_z=0$. The convention is to label states of
$J_z=0,1,2$ by $ \Sigma, \Pi, \Delta$  respectively. The ground state
($\Sigma^+_g$) will have $J_z=0$ and $CP=+$.

 The exploration of the energy levels  of other representations has a
long history in lattice studies~\cite{liv,pm}. The first excited state
is found  to be the $\Pi_u$.  This can be visualised  as the symmetry of
a string bowed out in the $x$ direction minus the same  deflection in
the $-x$ direction (plus another component of  the two-dimensional
representation with the transverse direction $x$ replaced by $y$),
corresponding to flux  states from a lattice  operator which is the
difference of U-shaped paths from quark to antiquark of the form $\,
\sqcap - \sqcup$. 

The picture of the gluon flux between the static  quarks suggests that
the excited states of this string may approximate the excited potentials
found from the lattice. In the simplest string  theory, the first
excited level has $\Pi_u$ symmetry and is at energy $\pi/R$ above the
ground state.  This is indeed  approximately valid and a closer
approximation  is to use a relativistic version~\cite{phm} (namely
$E_m(R)=(\sigma^2 R^2+2\pi\sigma (m-1/12))^{1/2}$ for the $m$-th level),
see also ref.~\cite{jkm} for a recent comparison of this expression.

Recent lattice studies~\cite{jkm}  have used an asymmetric space/time
spacing which enables excited states to be  determined comprehensively.
 These results confirm the finding that 
the $\Pi_u$ excitation is the lowest lying and hence of most relevance 
to spectroscopy.

 From the potential corresponding to these excited gluonic states, one
can  determine the spectrum of hybrid quarkonia using the Schr\"odinger
equation in the Born-Oppenheimer approximation.  This approximation will
be good if the heavy quarks move very little in the  time it takes for
the potential between them to become established. More  quantitatively,
we require that the potential energy of gluonic excitation is much
larger than the typical energy of orbital or radial excitation.  This is
indeed the case~\cite{liv}, especially for $b$ quarks. Another nice
feature of this approach is that the  self energy of the static sources
cancels in the energy difference between this  hybrid state and the
$Q \bar{Q}$ states. Thus the lattice approach gives directly the
excitation energy  of each gluonic excitation.

  The $\Pi_u$ symmetry state corresponds to  excitations of the gluonic
field in quarkonium called magnetic (with $L^{PC}=1^{+-}$) and
pseudo-electric (with $1^{-+}$) in contrast to the usual  P-wave orbital
excitation which has $L^{PC}=1^{--}$. Thus we expect different quantum
number assignments from those of the gluonic ground state. Indeed
combining with the heavy quark spins, we get a degenerate  set of 8
states:

\begin{table}
\begin{center}
\begin{tabular}{|c|c|ccc|} \hline
  $L^{PC}$ &   $J^{PC}$ &$J^{PC}$ &$J^{PC}$ &$J^{PC}$ \\ \hline
$1^{-+}$ & $1^{--}$ & $0^{-+}$ & $1^{-+}$ & $ 2^{-+}$ \\ 
$1^{--}$ & $1^{++}$ & $0^{+-}$ & $1^{+-}$ & $ 2^{+-}$ \\ 
\hline
\end{tabular}
\end{center}
\end{table}

\noindent  Note that of these,  $J^{PC}=  1^{-+},\ 0^{+-}$ and  
$2^{+-}$  are spin-exotic and hence will not mix with $Q\bar{Q}$ states.
They thus form a very attractive goal for experimental searches for
hybrid  mesons.

 The eightfold degeneracy of the static approach will be broken by 
various corrections. As an example, one of the eight degenerate  hybrid
states is a pseudoscalar with the heavy quarks in a spin triplet.  This
has the same overall quantum numbers as the S-wave  $Q \bar{Q}$ state
($\eta_b$) which, however, has the heavy quarks in a spin singlet. So
any  mixing between these states must be mediated by spin dependent
interactions.  These spin dependent interactions will be smaller for
heavier quarks. It is  of interest to establish the strength of these
effects for $b$ and $c$ quarks. Another topic of interest is the
splitting  between the spin exotic hybrids which will come from the
different  energies  of the magnetic and pseudo-electric gluonic
excitations.

 One way to go beyond the static approach is to use the NRQCD
approximation which then enables  the spin dependent effects to be
explored.  One study~\cite{jkm} finds that the  $L^{PC}=1^{+-}$ and
$1^{-+}$ excitations  have no statistically significant splitting 
although the $1^{+-}$  excitation does lie a little lighter. This would
imply, after adding in heavy quark spin, that  the $J^{PC}=1^{-+}$
hybrid was the lightest spin exotic. Also a relatively large spin
splitting was found~\cite{cppacs} among the triplet states considering,
however,   only  magnetic gluonic excitations.
 Another study~\cite{hyb-mix} explores the mixing of non spin-exotic
hybrids  with regular quarkonium states via a spin-flip interaction 
using lattice NRQCD.

\begin{figure}[th]

\epsfxsize=10cm\epsfbox{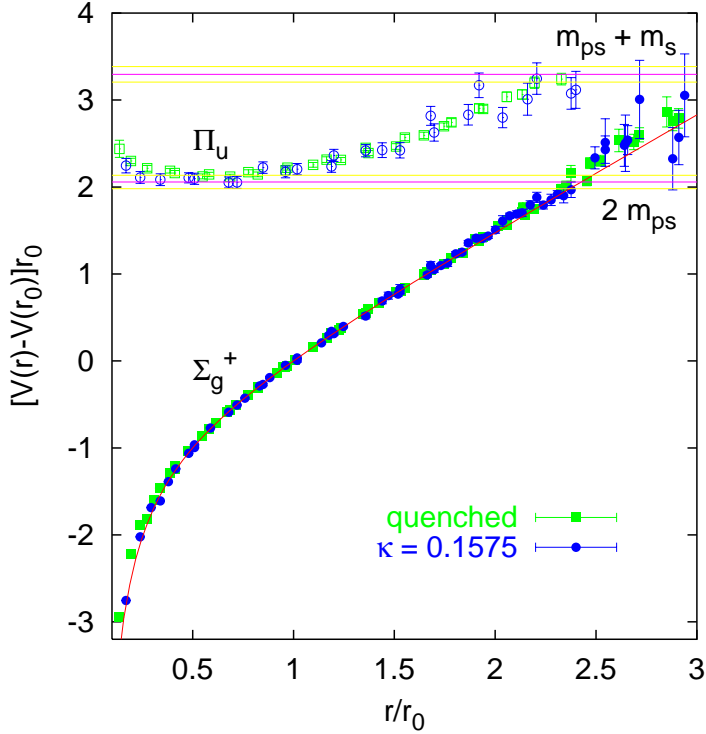}
 \caption{The potential energy for quenched and 2 flavours of sea quark
for the ground state and first excited gluonic
state~{\protect\cite{sesam}}.
 }
 \label{ex.balif}
\end{figure}

 Confirmation of the ordering of the spin exotic states also comes from
 lattice studies with propagating quarks~\cite{livhyb,milc,sesamhyb}
which  are able to measure masses for all 8 states. We  discuss that
evidence in more detail below.

 Because of the similarity of the lightest hybrid wavefunction with 
that of the 2S state (which has $L=1$), it is convenient to 
quote mass differences between these states.  
Within the quenched approximation,  the lattice evidence  for
$b\bar{b}$ quarks points to a  lightest hybrid spin exotic with
$J^{PC}=1^{-+}$ at an energy given by $(m_H-m_{2S})r_0$ =1.8 (static
potential~\cite{pm}); 1.9 (static potential~\cite{jkm},
NRQCD~\cite{cppacs}); 2.0 (NRQCD~\cite{jkm}). These results can be
summarised as       $(m_H-m_{2S})r_0=1.9 \pm 0.1$. 
 Using the experimental mass of the $\Upsilon(2S)$, this implies that
the lightest spin exotic  hybrid is at $m_H=10.73(7)$ GeV including a
10\% scale error.  Above this energy there will be many more hybrid 
states, many of which will be spin exotic.

 The results from a study  with $N_f=2$ flavours of sea-quarks show very
little change in the static potential (see  fig.~\ref{ex.balif}) and
relatively little change in NRQCD determinations~\cite{cppacs} of mass
ratios such  as $(m_H-m_{2S})/(m_{1P}-m_{1S})$.
 Expressed in terms of $r_0$  (using $r_0=1.18/\sqrt{\sigma}$) this gives
$(m_H-m_{2S})r_0=2.4(2)$, however.  This is significantly larger than
the quenched result and, using the $1P-1S$  mass difference to set the
scale, yields a prediction~\cite{cppacs} for the lightest hybrid mass  of
11.02(18) GeV.

\subsection{Hybrid meson decays}
 \label{ex.sect3.2}

 Within this static quark framework, one can explore the decay
mechanisms.  One special feature is that the symmetries of the quark and
colour fields about the static quarks must be preserved exactly in
decay.  This has the consequence that the decay from a $\Pi_u$ hybrid
state to the open-$b$ mesons ($B \bar{B},\   B^* \bar{B},\  B
\bar{B^*},\ B^* \bar{B^*}$) will be forbidden~\cite{hdecay} if the 
light quarks in the $B$ and $B^*$ mesons are in an S-wave relative to
the heavy quark (since the final state will have the light quarks in
either a triplet  with the wrong $CP$ or a singlet with the wrong $J_z$
where $z$ is the interquark axis). The decay to $B^{**}$-mesons with
light quarks in a P-wave is allowed by symmetry but not energetically.

The only allowed decays are when the hybrid state de-exites to a 
non-hybrid state with the emission of a light quark-antiquark pair.
Since the  $\Pi_u$ hybrid state has the heavy quark-antiquark in a
triplet P-wave state,  the resulting non-hybrid state must also  be in a
triplet P-wave since the heavy  quarks do not change their state in the
limit of very heavy quarks. Thus the decay for $b$ quarks will be  to
$\chi_b +M$ where $M$ is  a light quark-antiquark meson in a flavour
singlet. This proceeds by a disconnected  light quark diagram and it
would be expected~\cite{vall} that the scalar or pseudoscalar meson 
channels are the most important (ie they have the largest relative
OZI-rule  violating contributions).  Lattice estimates~\cite{hdecay} of
these  transitions have been made and the dominant mode (with a width of
around 100 MeV) is found to  be with $M$ as a scalar meson, namely $H
\to \chi_b + f_0$.

 These estimates are in the static quark limit, in which the spin-exotic
and non spin-exotic hybrid  mesons are degenerate. For the latter,
however,  the interpretation of any observed states is less clear cut,
since they  could be conventional quark antiquark states. Moreover, the
non spin-exotic  hybrid mesons can mix directly (ie without emission of
any meson $M$) with conventional quark antiquark states once  one takes
into account corrections (of order $1/M_Q$) to the static approximation 
applicable for  heavy quarks with physical masses. 

 It is encouraging that the decay width comes out as relatively 
small, so that the  spin-exotic hybrid states should show up 
experimentally as sufficiently narrow resonances to be detectable.
 This decay analysis does not take into account heavy quark motion or
spin-flip  and these effects will be significantly more important for
charm quarks than for $b$-quarks.

\subsection{Light quark hybrid mesons}
 \label{ex.sect3.3}

 I now  focus on lattice results for hybrid mesons made from light
quarks using fully relativistic propagating quarks.  There will be no
mixing with $q \bar{q}$ mesons for  spin-exotic hybrid mesons  and these
are of special interest. The first study of this area was by the  UKQCD
Collaboration~\cite{livhyb} who used operators motivated by the  heavy
quark studies referred to above to study all 8 $J^{PC}$ values coming
from $L^{PC}=1^{+-}$ and $1^{-+}$ excitations. The  resulting mass
spectrum  gives the $J^{PC}=1^{-+}$ state as the lightest spin-exotic
state. Taking account of the systematic scale errors in the lattice
determination, a  mass of 2000(200) MeV is quoted for this hybrid meson
with $s \bar{s}$ light quarks. Although not directly measured, the
corresponding light quark hybrid meson would be expected to be around
120 MeV lighter.

A second lattice group has also evaluated hybrid meson spectra with
propagating quarks  from quenched lattices. They obtain~\cite{milc}
masses of the $1^{-+}$ state with statistical and various different
systematic errors of  1970(90)(300)~MeV,  2170(80)(100)(100)~MeV and
4390(80)(200)~MeV for $n \bar{n}$,  $s \bar{s}$ and $c \bar{c}$ quarks
respectively. For the  $0^{+-}$ spin-exotic state they have a noisier
signal but evidence that it is heavier. They also explore mixing matrix
elements between spin-exotic hybrid  states and 4 quark operators.

 The first analysis~\cite{sesamhyb} to determine the hybrid meson
spectrum using  full QCD used Wilson quarks. The sea quarks used had
several different masses and an extrapolation  was made to the limit of
physical sea quark masses, yielding a mass of 1.9(2) GeV for the
lightest  spin-exotic hybrid meson, which again was found to be the
$1^{-+}$. In principle this  calculation should take account of sea
quark effects such as the mixing  between such a hybrid meson and $q
\bar{q} q \bar{q}$ states such as $\eta \pi$, although it is possible
that the sea quark  masses used are not light enough to explore these
features.

 A  recent dynamical quark study from 2+1 flavours of improved staggered
quarks has also produced results~\cite{milc2}. They also compare their
results with quenched calculations and find  no significant difference,
except that the ambiguity in fixing the lattice  energy scale is better
controlled in the dynamical simulation since different reference
observables are closer to experiment. Their summary result for the
$1^{-+}$ hybrid with strange quarks is  $2100 \pm 120$ MeV, in agreement
with earlier results. They note that the  energies of  two-meson states
(such as $\pi + b_1$ or $K + K(1^{+})$ ) with the hybrid  meson quantum
numbers are close to the energies they obtain. This suggests that  these
two-particle states, which are allowed to mix in a dynamical quark
treatment, may be  influencing the masses determined. A study of hybrid
meson transitions to two particle states is needed  to illuminate this
area, using techniques such as those used for heavy quark hybrid
decay~\cite{hdecay}  and  decays of light quark  vector 
mesons~\cite{rhodecay}.

The lattice calculations~\cite{livhyb,milc,sesamhyb,milc2,ml} of the
light hybrid spectrum are  in good agreement with each other. They imply
that the natural energy  range for spin-exotic hybrid mesons is around
1.9 GeV. The $J^{PC}=1^{-+}$  state is found to be lightest. It is not
easy to reconcile these lattice results  with experimental
indications~\cite{expt} for resonances at 1.4 GeV and 1.6 GeV,
especially the  lower mass value.  Mixing  with  $q \bar{q} q \bar{q}$
states such as $\eta \pi$ is not included for realistic quark masses in
the  lattice calculations. Such effects of pion loops (both real and
virtual) have been estimated  in chiral perturbation theory based
models~\cite{aa} and they could potentially reconcile some of the
discrepancy between lattice mass estimates (with light quarks which are
too heavy)  and those from experiment. This can be interpreted,
dependent on one's viewpoint,  as either that the lattice calculations 
are incomplete or as an indication that the experimental states may have
an  important meson-meson component in them.

 The light quark technique of using relativistic propagating quarks 
can also be extended to charm quarks, as was note above~\cite{milc}.
 Another group has explored the charm quark hybrid states also using a
fully  relativistic action, albeit with an anisotropic  lattice
formulation~\cite{manke}. Their quenched study is in agreement with the
isotropic lattice result quoted above, finding  a mass value of 
4.428(41) GeV in the continuum limit for the  $1^{-+}$ hybrid where the
scale is set by the $^1P_1 - 1S$ mass splitting (458.2 MeV
experimentally) in charmonium.  Their result is also consistent with
that from NRQCD methods~\cite{cppacs} applied to this case. These 
results all have the usual caveat that in quenched evaluations the
overall mass scale of the energy difference  from the $1S$ state at
3.067 GeV is uncertain to  10\% or so (for example the
$(2S-1S)/(^1P_1-1S)$ is  found to be 15\% higher than experiment) which
is a major source of systematic error (approximately $\pm 140$ MeV). 
They  also produce estimates for other charmonium spin-exotic states: 
$0^{+-}$ at 4.70(17) GeV and $2^{+-}$ at 4.89(9) GeV. The $0^{--}$ state
is not resolved.

Thus  masses near 4.4GeV  are found for the  charmonium $1^{-+}$ state
using relativistic quarks. The non-relativistic approach using  NRQCD is
expected to have big systematic errors for quarks as light as charm, but
results~\cite{cppacs} do agree with this value.
 The heavy quark effective theory approach has a leading term which
corresponds to a static heavy quark,  resulting in an estimate~\cite{pm}
of  the spin-exotic charm  state mass of 4.0 GeV.  Here again the 
systematic error is potentially large for charm quarks.

\section{Hadronic molecules}
 \label{ex.sect4}

 By exotic state we mean any state which is not dominantly a $q \bar{q}$
 or $qqq$ state. For example, a state made from hadrons bound in a molecule 
would be exotic.  
 Examples of hadronic molecules have been known for a long time: the
deuteron  is a proton-neutron molecule for example. It is very weakly
bound (2 MeV)  and is quite extended. It is more efficiently described 
in terms to a neutron and a proton than as six quarks.

 The residual hadronic interaction, the force between two colour-singlet
hadrons, is much weaker than the colour force between quarks. Although
it is called a \lq strong interaction\rq, it is relatively weak. At
large  distance, it will be dominated by the exchange of the lightest
hadrons  allowed (typically one or two pions). For example, the scale of
nuclear binding is around 8 MeV whereas the  gluonic forces binding
three quarks to make a nucleon contribute most of its  mass of 938 MeV.
For this reason lattice methods need to be developed  specially to
tackle this problem. Basically, one is interested in binding energies, 
so it is the energy difference between the two hadrons and the hadronic
molecule that is of interest.  This difference can sometimes  be
determined better than the total energy itself. Even so, the detailed
dynamics of such molecular states will depend  on the long range forces
(typically one or two pion exchange) and this will be  modified
considerably in lattice studies with light quark masses which are too
heavy (typically  down to 50\% of the strange quark mass only). So only
qualitative input  can be obtained from the lattice, but this can still
be used to validate   models.

 Because of the small  binding energy and the dominance of pion exchange
in the binding, it is not feasible  to obtain the deuteron binding
direct from QCD using lattice methods at present. 
  There has been speculation that other di-baryon systems might be more
strongly bound: the  H dibaryon (a $\Lambda \Lambda$ state) being the
best known. If it were strongly bound then it might be stable to weak
decay which might have  astrophysical consequences. The current status
of lattice studies~\cite{hdibaryon} is that finite box size effects are
large but there is no convincing lattice evidence that this state is
bound. At the largest volumes studied, with $L \approx 4$fm,  in
quenched simulations  the ratio $(m_H /2 m_{\Lambda}) -1$ is found to be
positive  and in the range 5 to 15 \%.

 Other molecular states involving two hadrons have been conjectured.
Several  meson resonances are known which are closely connected with
nearby thresholds:  the $\Lambda(1405)$ which is just below the
$\bar{K}N$ threshold and  the $a_0(980)$ and $f_0(980)$  which are close
to the $K \bar{K}$ threshold. Another state close to a threshold is the 
$N(1535)$ which is just above the $\eta N$ threshold.  Again lattice
studies are not able to shed very much light directly on these states
since the  quark masses used in the lattice studies are unphysical.
However, if they are not produced in a lattice study which explores $q
\bar{q}$  and $qqq$ states, this may help to support the conclusion that
they  are primarily molecular in structure.

\begin{figure}[bt] 
\vspace{-2.0cm}  
\epsfxsize=10cm\epsfbox{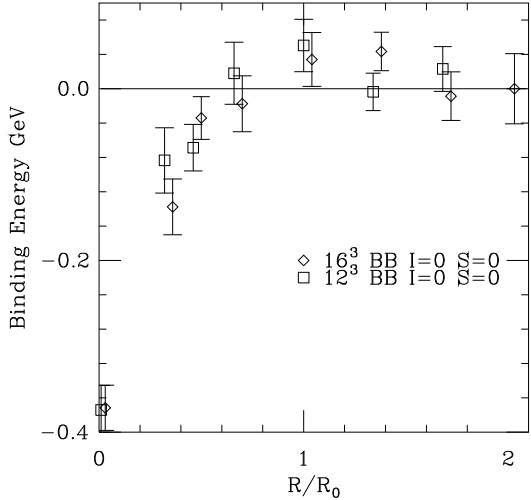}

 \caption{ The binding energy~{\protect\cite{cmpp}} between two
heavy-light mesons (with static heavy quarks and light quarks of mass
corresponding to strange) at separation $R$ (in units of  $r_0 \approx
0.5$ fm) with the two light quarks having  I=0 and S=0. 
   }
 \label{ex.bb00}
\end{figure}

\begin{figure}[bt] 
\vspace{-2.0cm}  
\epsfxsize=10cm\epsfbox{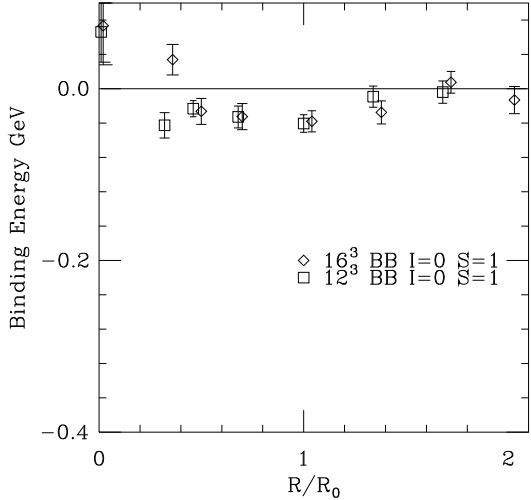}

 \caption{ The binding energy~{\protect\cite{cmpp}} between two
heavy-light mesons (with static heavy quarks and light quarks of mass
corresponding to strange) at separation $R$ (in units of  $r_0 \approx
0.5$ fm) with the two light quarks having  I=0 and S=1. 
   }
 \label{ex.bb01}
\end{figure}

 One case which is relatively easy to study is the $BB$ system,
idealised as two  static quarks and two light quarks. Then a potential
as a function of the separation  $R$ between the static quarks can be
determined.  Because the static quark spin is irrelevant, the states can
be classified by the light quark spin and  isospin.  Lattice
results~\cite{cmpp}  (using a light quark mass close to strange) have
been obtained for the potential energy for $I_q=0,1$ and $S_q=0,1$. For
very  heavy quarks, a potential below $2M_B$  will imply binding of the
${BB}$ molecules with these quantum numbers and $L=0$. For the
physically relevant case  of $b$ quarks of around 5 GeV, the kinetic
energy will not be negligible and the binding energy of the ${ BB}$
molecular states is less  clear cut. One way to estimate the kinetic
energy for the ${ BB}$ case with reduced mass circa 2.5 GeV is to use
analytic approximations to the  potentials found. For example the
$I_q,S_q$=(0,0) case (see fig.~\ref{ex.bb00}) shows a deep  binding at
$R=0$ which  can be approximated as a Coulomb potential of $-0.1/R$ in
GeV units. This will give a di-meson binding energy of only 10 MeV.  For
the other interesting case shown in fig.~\ref{ex.bb01},
$(I_q,S_q)$=(0,1), a  harmonic oscillator potential in the radial
coordinate of form $-0.04[ 1- (r-3)^2/4]$ in GeV units leads to a
kinetic energy  which completely cancels the potential energy minimum,
leaving zero  binding. This harmonic oscillator approximation lies above
the estimate of  the potential, so again we expect weak binding of the
di-meson system.

 Because of these very small values for the di-meson binding energies, 
one needs to retain corrections to the heavy quark approximation to 
make more definite predictions, since these corrections are known to 
be of magnitude 46 MeV from the $B$, $B^*$ splitting. It will also be 
necessary to extrapolate the  light quark mass from strange to 
the lighter $u,\ d$ values to make more definite predictions 
about the binding of $BB$ molecules.

\begin{figure}[bt] 
\vspace{-2.0cm}  
\epsfxsize=9cm\epsfbox{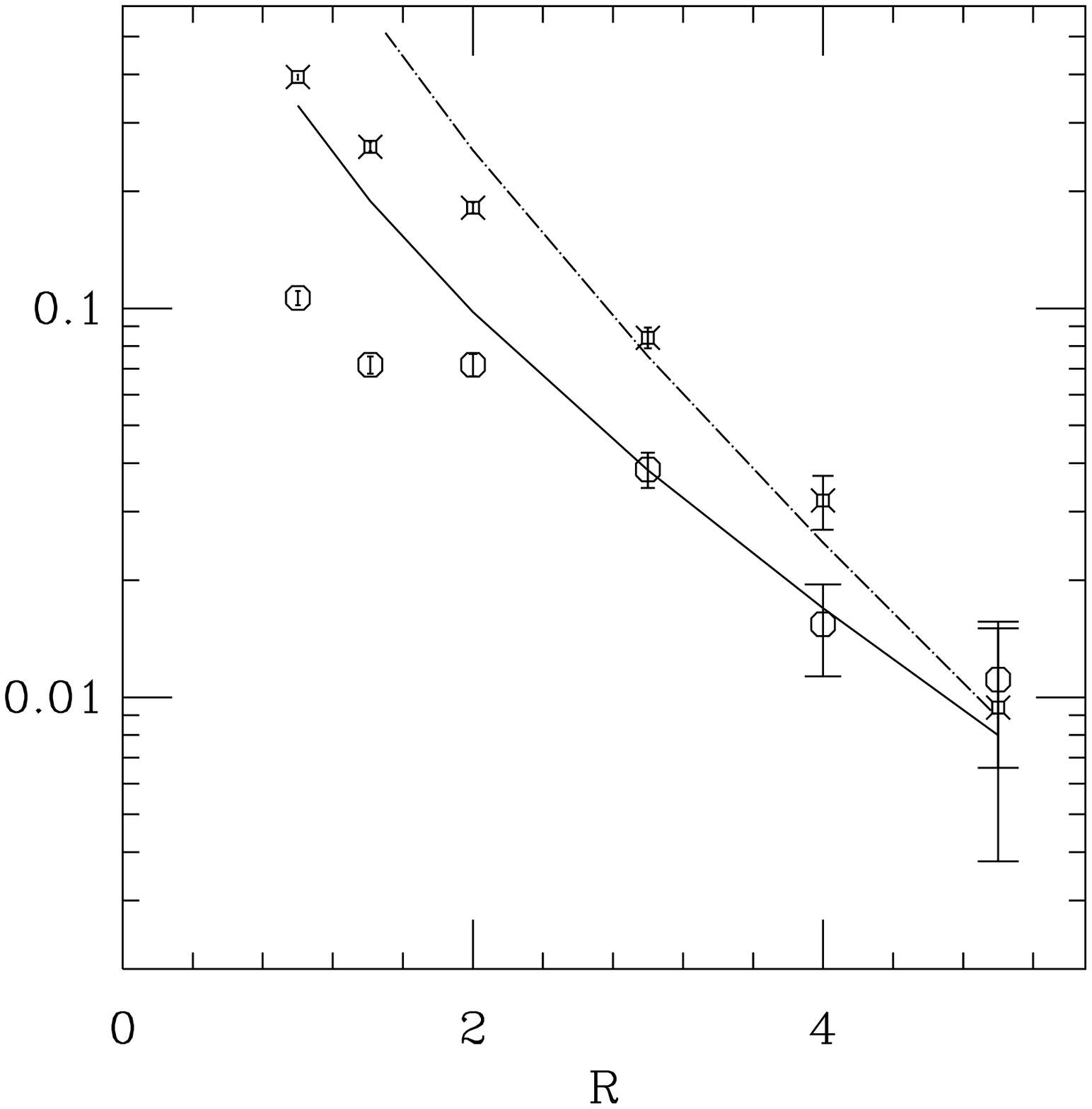}

 \caption{ The contribution~{\protect\cite{cmpp}}  to the binding energy
for the spin and isospin combinations  corresponding to $\pi$ exchange
(octagons) and $\rho$ exchange (fancy squares). The solid line gives the
$\pi$ exchange contribution which is normalised by the  $B^{*} B \pi$
coupling. The $\rho$ exchange prediction has a free normalisation  and
is shown by the dash-dotted line. The results are plotted versus
interquark separation R (in units of  $a \approx 0.17$ fm).
   }
 \label{ex.pirho}
\end{figure}

Models for the binding of two $B$ mesons involve, as in the case of the
deuteron,  pion exchange. The lattice study~\cite{cmpp} is able to make
a quantitative comparison of lattice pion  exchange with the data
described above using lattice determinations of  the $B^{*}B\pi$
coupling~\cite{bbpi} and excellent agreement is obtained at larger $R$ 
values as shown in fig.~\ref{ex.pirho}, as expected.

\section{Conclusions and Outlook}
 \label{ex.sect5}

 Quenched lattice QCD is well understood and accurate predictions in the
continuum limit  are increasingly becoming available. The lightest
glueball  is scalar with mass  $m(0^{++})=1611(30)(160)$ MeV where the
second error is an overall scale error. The excited glueball spectrum is
known too. The quenched approximation  also gives information on
quark-antiquark scalar mesons and their mixing with glueballs. This
determination of the mixing in the quenched approximation  also sheds
light on results for the  spectrum directly  in full QCD where the
mixing will be enabled. In full QCD, the scalar meson masses are
determined directly but there is no concept of a glueball as such, much
as  in the experimental case. Additional work is need to reduce the
lattice spacing, or use improved actions,  to explore the continuum
limit for scalar mesons in full QCD. 
  There is also some lattice information  on the hadronic decay 
amplitudes of glueballs and this is an area where further study may be
anticipated.

 For hybrid mesons, there will be no mixing with $q \bar{q}$ for 
spin-exotic states and these are the most useful predictions. The
$J^{PC}=1^{-+}$ state is expected in the range 10.7 to 11.0 GeV for $b$ quarks,
 2.0(2) GeV for $s$ quarks and 1.9(2) GeV 
for $u,\ d$ quarks. Mixing of spin-exotic hybrids with
$q\bar{q}q\bar{q}$ or equivalently with meson-meson  is  allowed and
will modify the  predictions from the quenched approximation.
 A first lattice study has been made of hybrid meson decays. For heavy 
quarks, the dominant mode is string de-excitation to $\chi + f_0$ where
$f_0$ is  a flavour singlet scalar meson (or possibly two pions in this
state). The magnitude of the decay rate is found to be of order 100 MeV, 
so this decay mode should  still leave a detectably narrow resonance to 
be observed. 

 The topic of possible multi-quark bound states is difficult because the
 scale of the expected binding energies is a few MeV and this small
value is  a challenge  for lattice studies.  As an example, some
evidence was presented  for a  possible $B_s B_s$ molecular state.


\end{document}